\documentclass[lettersize,journal]{IEEEtran}
\usepackage{amsmath,amsfonts}
\usepackage{algorithmic}
\usepackage{array}
\usepackage[caption=false,font=normalsize,labelfont=sf,textfont=sf]{subfig}
\usepackage{textcomp}
\usepackage{stfloats}
\usepackage{url}
\usepackage{verbatim}
\usepackage{graphicx}
\usepackage{cite}
\usepackage{multirow}
\usepackage[ruled,lined,boxed,commentsnumbered]{algorithm2e}
\usepackage{color}   
\usepackage{xcolor}
\usepackage{xpatch}

\begin{document}

\title{Mobile Recording Device Recognition Based Cross-Scale and Multi-Level Representation Learning}

\author{Chunyan~Zeng, Yuhao~Zhao, Zhifeng~Wang

\thanks{Chunyan~Zeng and Yuhao~Zhao are with the Hubei Key Laboratory for High-efficiency Utilization of Solar Energy and Operation Control of Energy Storage System, Hubei University of Technology, Wuhan 430068, China.}
\thanks{Zhifeng~Wang is with the Department of Digital Media Technology, Central China Normal University, Wuhan 430079, China (Corresponding author: Zhifeng Wang, E-mail: zfwang@ccnu.edu.cn).}
}

\markboth{arxiv}%
{Zeng \MakeLowercase{\textit{et al.}}}


\maketitle

\begin{abstract}
The issue of identifying the source of mobile recording devices is a crucial focus in the realm of consumer electronics applications. It significantly aids in evidence collection for judicial authorities and the protection of individual intellectual property.
Existing studies predominantly utilize a single network module to deeply analyze input features at the frame-level, neglecting the representation of audio at the sample-level and the integration of global temporal information. To address these gaps, this paper introduces a modeling approach that employs multi-level global processing, encompassing both short-term frame-level and long-term sample-level feature scales.
In the initial stage of shallow feature extraction, various scales are employed to extract multi-level features, including Mel-Frequency Cepstral Coefficients (MFCC) and pre-Fbank log energy spectrum.
The construction of the identification network model involves considering the input two-dimensional temporal features from both frame and sample levels. Specifically, the model initially employs one-dimensional convolution-based Convolutional Long Short-Term Memory (ConvLSTM) to fuse spatiotemporal information and extract short-term frame-level features.
Subsequently, bidirectional long Short-Term Memory (BiLSTM) is utilized to learn long-term sample-level sequential representations.
The transformer encoder then performs cross-scale, multi-level processing on global frame-level and sample-level features, facilitating deep feature representation and fusion at both levels.
Finally, recognition results are obtained through Softmax.
Our method achieves an impressive 99.6\% recognition accuracy on the CCNU\_Mobile dataset, exhibiting a notable improvement of 2\% to 12\% compared to the baseline system. Additionally, we thoroughly investigate the transferability of our model, achieving an 87.9\% accuracy in a classification task on a new dataset.
	
\end{abstract}


\section{Introduction}
\IEEEPARstart{T}{he} pervasive use of digital media, including audio \cite{Wang2011a} and video \cite{Tian2018,Wang2021}, in fundamental social contexts such as communication, socialization, and media has significantly simplified information dissemination in consumer electronics applications \cite{Prabhakar2023,Kawase2020,Chen2023b}. However, the ease of editing and modifying digital media files, facilitated by various software applications \cite{DosSantos2023,bib26}, has also made it remarkably convenient for malicious entities to tamper with and forge audio and video files \cite{Wang2022t}, raising concerns about information security and escalating associated risks \cite{bib1,Zeng2022a,bib2,Zeng2023a,bib3}.

To counter these challenges, the emergence of mobile recording device identification technology for digital media audio files has proven vital \cite{Zeng2024a}. Prior research confirms that each digital audio recording device introduces a unique "machine fingerprint" during the recording process, stemming from subtle differences in hardware and software configurations \cite{Chen2023c,Zeng2024b}. The key to solving the mobile recording device identification problem lies in extracting representative feature information from recorded audio to ensure heightened sensitivity in recognizing these machine fingerprints \cite{Wang2023f}. This paper explores strategies for extracting finer and more discernible features from audio data to enhance machine fingerprint identification accuracy.

Research on mobile recording device source identification comprises two main components: frontend feature extraction and backend recognition model construction. For the frontend, two primary methods of feature extraction exist. The first method focuses on spectral features of the original audio signal, such as MFCC \cite{bib4,bib5,Zeng2024d} and Band Energy Difference (BED) \cite{bib6}. These methods utilize acoustically relevant prior knowledge to extract spectral features or other distinguishing features and employ techniques like Voice Activity Detection (VAD) \cite{bib7} and spectral subtraction \cite{bib8} to enhance recognition accuracy. The second method is based on the representation of key information in the probability model of the original audio. Researchers use MFCC to construct Gaussian Mixture Models (GMM) and extract Gaussian Supervector Vectors (GSV) \cite{bib9} to represent audio features, often combined with backend work in classical machine learning.

Backend recognition models fall into two categories: traditional machine learning methods, such as Support Vector Machine (SVM) \cite{bib4,bib10} and GMM\cite{bib11,bib12,bib13}, and representation learning based on neural network models, incorporating deep learning technologies like Bidirectional Encoder Representations from Transformers (BERT) and Transformer from other domains \cite{Zheng2024}. 

Despite the application of deep neural networks, current research predominantly focuses on frame-level features of audio data, with limited consideration for sample-level features on a global scale. This paper emphasizes the extraction of audio features at multiple levels and explores temporal and spatial features locally and globally in the backend recognition model to improve recognition performance. The main contributions of this paper can be summarized as follows:

\begin{itemize}
	\item In the frontend feature extraction, this paper focuses on the frequency domain features of original audio, conducting multiple shallow feature extractions and fusions. The extracted MFCC features and pre-Fbank features are fused based on time and space representations.
	\item In the backend recognition network, this paper conducts temporal and spatial representation learning of frame-level and sample-level features, thoroughly mining device source information embedded in shallow features from multiple scales. Through ConvLSTM and BiLSTM networks, information mining is performed on frame-level and sample-level scales, obtaining embeddings with a larger receptive field.
	\item In the backend recognition network, this paper also conducts representation learning on global features, interacting information from various scales (frame-level, sample-level, and cross-scale level). Through Transformer-encoder, deeply processed depth information at multiple levels is globally processed, providing better representation for embeddings and benefiting the final recognition performance.
	\item In experimental results, the proposed method in this paper exhibits precise recognition effects on the CCNU\_Mobile dataset, achieving numerical values of 99.6\% in recognition accuracy, recall, precision, and F1-Score. This proves the effectiveness of the method proposed in this paper. Furthermore, through fine-tuning pre-trained models and utilizing small batch sample transfer training, the method achieves a recognition accuracy of 87.9\% on a new database.
\end{itemize}

The structure of the remaining sections of this paper is as follows: Section \ref{RW} extensively reviews existing methods and related work in the literature. Section \ref{Pre} initiates a preliminary discussion, defines the problems to be addressed, and provides theoretical support, including a list of relevant symbols, for reader reference. Section \ref{4} provides a detailed introduction to the principles and theoretical support of the proposed method. Section \ref{Exp} introduces experiments designed to prove the effectiveness and superiority of the method through analysis of experimental data. Finally, Section \ref{Con} summarizes this study and conducts a comprehensive discussion of identified shortcomings.


\section{Related Work} \label{RW}

In this section, we delve into existing research in both frontend and backend aspects. In the frontend domain, we specifically highlight frequency domain representation, closely aligned with this paper's focus. Meanwhile, in the backend domain, we categorize the exploration into two realms: classical machine learning and deep learning.

\subsection{Information characterization of recording devices based on spectral feature engineering}
The evolution of acoustic features has significantly propelled advancements in the field of recording device source recognition. This involves a comprehensive analysis of various aspects of audio signals to derive distinctive feature representations. These representations encompass spectrum, phase inverted spectrum, power, dynamic range, peak factor, and time-varying autocorrelation. Depending on the adopted signal analysis methods, the representation of recording device information can be broadly classified into two categories: frequency spectral class features and cepstrum class features

Regarding frequency spectral class features, Buchholz et al. \cite{bib18} introduced the use of the Fourier coefficient histogram as a feature, pioneering the application of frequency domain features obtained through short-time Fourier transform for recording device source identification. To enhance the representativeness and generalization of these frequency domain features and optimize the feature extraction process, researchers explored methods to augment information through feature mapping. Panagakis et al. \cite{bib19,bib20} proposed two types of frequency domain features, namely Random Spectral Features (RSF) and Labeled Spectral Features (LSF), through unsupervised and supervised learning, respectively. Luo et al. \cite{bib6} presented BED features, demonstrating the effectiveness of extracting frequency response curves from sample recordings. The energy difference feature extracted could reflect temporal patterns in audio signals, showcasing significant discriminative capabilities in recording device recognition tasks. Despite the effectiveness of these spectrum features, their extraction relies on intricate mappings, simultaneously increasing computational workload and complexity while enhancing feature representation. Additionally, finding more effective methods to eliminate interference information during the feature extraction process remains a challenge.

Cepstrum class features are obtained by logarithmically processing the spectrum and then performing inverse Fourier transform. Currently, they find widespread application in research on recording device source recognition. Kraetzer et al. \cite{bib21} classified four recording devices based on time, frequency, and cepstral domain features, proposing the use of Mel frequency cepstral coefficients as machine fingerprints for recording device identification. Building upon this, Hanilci et al. \cite{bib7} utilized Voice Activity Detection (VAD) to detect speech segment endpoints, extracting MFCC from non-speech segments. They verified that features extracted from non-speech segments of audio could, to some extent, eliminate interference from speech information. However, due to audio data being primarily composed of speech segments, isolating sufficiently long silent segments for feature extraction poses a challenge in practical applications. Aggarwal et al. \cite{bib8} separated estimated noise signals from speech using spectral subtraction, obtaining noise spectrum signals. This method, compared to using original MFCC features, improved recognition accuracy. In addition to MFCC, other cepstrum features such as Linear Frequency Cepstral Coefficients (LFCC) and Linear Prediction Cepstral Coefficients (LPCC) are also applied in recording device recognition. Hanilci et al. \cite{bib22}, focusing on cepstral features, employed Cepstral Mean and Variance Normalization (CMVN) to standardize features, enhancing the robustness of cepstral features for recording device recognition.

\textbf{Summary:} These acoustic feature extraction methods lay a solid foundation for accurate recording device recognition. Moreover, these methods extend to other fields such as speaker recognition and emotion recognition. However, in the recording device recognition scenario, a more targeted feature description scheme is needed to further improve the application of these features in specific environments.

\subsection{Recognition of Recording Devices Based on Classical Machine Learning Algorithms}

This category of methods typically employs extracted acoustic features as inputs for machine learning algorithms to identify the source device of the recording. Backend classification algorithms mainly fall into three categories: recording device recognition models based on SVM, GMM, and Sparse Representation Classifier (SRC).

SVM performs nonlinear classification by dividing data in the mapped space, adjusting decision boundaries under supervised learning, and using a learning strategy that maximizes inter-class intervals. Hanilci et al. \cite{bib4} used MFCC as input features, employed the Vector Quantization (VQ) algorithm, and SVM as the backend classification decision. They analyzed the spectral information differences of different brands of devices, verifying the effectiveness of SVM as a recognition algorithm in recording source device recognition tasks. Subsequently, SVM, as a benchmark classification model, has been widely used in the field of recording device source recognition \cite{bib17,bib23,bib24}.

GMM models can precisely represent the attributes of things through a probability density model \cite{bib13,bib19}. Early research on GMM involved training GMMs for each device using multiple Gaussian mixture components, then determining the GMM that best matches the audio sample information through maximum likelihood estimation, achieving the purpose of classifying recording devices \cite{bib7,bib12,bib25}. Kotropoulos et al. \cite{bib10} utilized GMM to extract more characteristic recording device features, inputting features such as MFCC and LFCC into GMM to construct GSV. They achieved high recognition accuracy through classification algorithms such as SVM and Radial Basis Function-Neural Network (RBF-NN). GMM-based methods have achieved certain research results in the field of recording device recognition, especially GSV-type features that efficiently reflect device-related information in feature data. However, in the GMM training process, choosing the appropriate number of Gaussian mixture components and designing the appropriate parameter training iteration process may lead to a significant increase in model complexity.

Recording device recognition models based on SRC classify by constructing a complete function dictionary and sparse representation feature data matrix, reducing interference information in features, and retaining key information in channel mode noise. Zou et al. \cite{bib12} built a supervised learning dictionary using GSV and MFCC, comparing and calculating the distance differences between sparse matrices to find suitable sample attribute classifications. Moreover, they used similarity matching based on KISS metric learning to record device information for sparse representations.

\textbf{Summary:} Traditional machine learning methods distinguish and classify input feature data by mapping it to a higher-dimensional space, deeply exploring the recognition of recording device sources from the perspective of information representation. However, traditional supervised machine learning algorithms lack the ability to deeply represent data in recording device recognition research.

\subsection{Recognition of recording device based on deep learning technique}

The rapid evolution of neural networks and deep learning technologies in recent years has spurred researchers to incorporate these methods from diverse domains into the realm of recording device recognition. Building precise recognition network models proves advantageous in extracting more representative deep features in audio \cite{Zhu2013}.

Li et al. \cite{bib14,bib15} introduced a method utilizing MFCC features for training a deep autoencoder network, followed by using the output from the bottleneck layer as the final feature representation. Verma et al. \cite{bib27} conducted a comparison between Discrete Fourier Transform (DFT) frequency features and BED features on a designed Convolutional Neural Network (CNN) model. Lin et al. \cite{bib16} proposed a sub-band perception self-attention mechanism CNN, directing attention to the most relevant frequency band segments in the spectrogram for a more efficient feature representation. This model achieved remarkable recognition accuracy in both noise-free and noisy scenarios. Shen et al. \cite{bib28} presented a recording device recognition model based on the Residual Neural Network (ResNet), combining ResNet with GoogLeNet to construct a comprehensive network model. VAD and spectral subtraction were employed during feature extraction for effective noise reduction. Zeng et al. \cite{bib29} put forth a multi-feature fusion recording device recognition method, utilizing deep representation learning to extract crucial information features for recording device recognition. They also proposed a spatiotemporal representation learning model based on an attention mechanism , resulting in elevated recognition accuracy. Naini et al. \cite{bib30} amalgamated CNN and Long Short-Term Memory (LSTM) networks, presenting a dual attention pooling network within the CNN-LSTM framework. Experimental results demonstrated the efficacy of this strategy in amalgamating the distinct advantages of the two networks, surpassing the recognition performance of a singular network.

\textbf{Summary:} Recording device recognition methods grounded in deep learning exhibit robust representation learning capabilities, automatically distilling highly condensed deep features from raw data. Despite their effectiveness, these models grapple with challenges such as insufficient prior information and limited alignment with the target task. Unstructured deep learning models, driven purely by data, lack a systematically designed structure that explicitly encapsulates the features relevant to recording device recognition. A thorough investigation into these challenges will contribute to the refinement and optimization of deep learning models in this domain.

\section{Preliminaries} \label{Pre}

To facilitate a thorough comprehension of the methods presented in this paper and improve overall readability, this section will employ a mathematical model to articulate the problem, as shown in Table \ref{tab:table1}. Furthermore, it will enumerate the relevant concepts, symbols, and definitions that will be employed in the subsequent chapters.
\begin{table}[h]
	\caption{A summary of notations.\label{tab:table1}}
	\centering
	\begin{tabular}{l|l}
		\hline
		\textbf{Notation}           & \textbf{Description}                                 \\ \hline
		$x(n)$                      & Speech Signal after Frame Segmentation and Windowing \\
		$f_F( \cdot )$              & Deep Learning Model for Frame-Level Feature Input    \\
		$f_S( \cdot )$              & Deep Learning Model for Sample-Level Feature Input   \\
		$f_C$                       & Multi-level Fusion Features for Device Integration   \\
		$C_{S}\left [ \cdot\right]$ & Concatenation Operation                              \\
		$*$                         & Convolution Operation                                \\
		$\odot$                     & Matrix Hadamard Product                              \\
		$\sigma (\cdot )$           & Sigmoid Activation Function                          \\
		$i_{t}$                     & Input Gate of ConvLSTM at Time Step $t$              \\
		$f_{t}$                     & Forget Gate of ConvLSTM at Time Step $t$             \\
		$o_{t}$                     & Output Gate of ConvLSTM at Time Step $t$             \\
		$g_{t}$                     & Input Modulation Gate of ConvLSTM at Time Step $t$   \\
		$x_{t}$                     & Input Data at Time Step $t$                          \\
		$c_{t}$                     & Cell State at Time Step $t$                          \\
		$h_{t}$                     & Hidden State at Time Step $t$                        \\
		$d_{k}$                     & Dimension of the Key Matrix $K$                      \\
		$head_{h}$                  & Attention Mechanism of the $h$-th Head               \\
		$W^{O}$                     & Weights Calculated During Global Training            \\
		$\mu$                       & Mean of Normalization Parameters                     \\
		$\sigma$                    & Variance of Normalization Parameters                 \\
		$H$                         & The Number of Hidden Units in One Layer              \\
        $TP_{i}$  & The TP value for the $i-th$ class sample as an example              \\
        $FP_{i}$ & The FP value for the $i-th$ class sample as an example             \\
        $TN_{i}$  & The TN value for the $i-th$ class sample as an example              \\
        $FN_{i}$  & The FN value for the $i-th$ class sample as an example              \\\hline
	\end{tabular}
\end{table} 

\subsection{Problem Formalization}

\begin{figure}[h]
\centering
\includegraphics[width=3.5in]{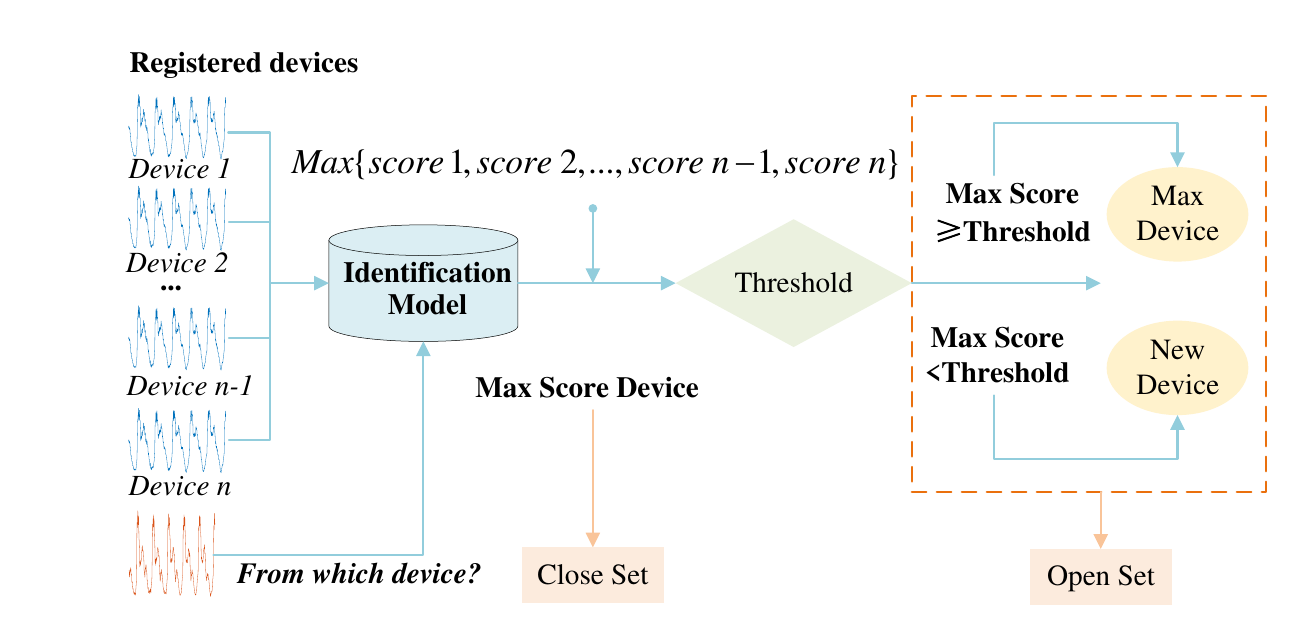}
\caption{Flowchart of mobile recording device recognition problem.}
\label{fig_1}
\end{figure}

\textbf{Problem Definition} (\textit{Mobile Recording Device Identification Problem}).
The primary objective of addressing the digital audio mobile recording device source identification problem is to construct a sophisticated model $\mathcal{M}(\cdot)$ using state-of-the-art techniques in the field. This model is designed to accurately attribute a given test audio $v^{t}$ to its source device within a registered database $\left \{ v_{n}^{e}|n=1,2,3,...,N  \right \}$. Depending on whether the test audio $v^{t}$ must belong to the registered database, the problem is classified as either a closed-set or an open-set problem. This process is mathematically described by Eq. \ref{eq1}:

\begin{equation}
\label{eq1}
n^{*}= \mathop{argmax}\limits_{n}\left \{ \mathcal{M}(v_{1}^{e},v^{t}),\mathcal{M}(v_{2}^{e},v^{t}),...,\mathcal{M}(v_{N}^{e},v^{t}) \right \}  
\end{equation}
Where $N$ denotes the total number of devices in the registered database, and $n^{*}$ represents the device identified in the final output. Each entry corresponds to the similarity score computed by the model for the test audio, and the device with the highest score is ultimately recognized. The device identification problem is visualized in the flowchart presented in Fig. \ref{fig_1}.

\subsection{Definitions}

\textbf{Definition 1} (\textit{Frame-level Feature Representation Learning}).
Frame-level feature representation learning refers to the process of segmenting the speech signal $s(n)$ into frames and utilizing a deep neural network to transform each frame (represented as $x(n)$) into a set of feature vectors denoted as $f_{F}$. This process embeds crucial information from the speech signal, including spectrum, energy, cepstrum, etc. To mitigate frame boundary discontinuity, a window function $\omega (n)$ is conventionally applied during the framing process. Various frame-level signals are obtained using filter banks at different scales, as depicted in Eq. \ref{eq2}.

\begin{equation}
\label{eq2}
x(n)=Fbank\left [ s(n)\times \omega (n) \right ]   
\end{equation}

The deep learning model $\mathcal{M}_{F}$ is trained to acquire frame-level feature representations $f_{F}$, aiming to enhance the comprehension and processing of the speech signal $s(n)$. This enhancement is intended to boost the overall performance of the task of identifying mobile recording devices, as depicted in Eq. \ref{eq3}.

\begin{equation}
\label{eq3}
f_{F} = \mathcal{M}_{F} \left\{ x(n) \right\}
\end{equation}



\textbf{Definition 2} (\textit{Sample-level Feature Representation Learning}). 
Speech signals are one-dimensional temporal signals and inherently exhibit temporal sequences when presented as inputs to the model, sample-level feature representation learning involves designing a deep model $\mathcal{M}_{S}$. This model aims to extract sample-level feature embeddings $f_S$ from multiple speech samples $\left [ ..., x(t-1), x(t), x(t+1), ... \right ]$ with temporal correlations. The primary goal is to employ deep learning to model speech signals, learning to integrate temporal information and yield a more representative feature representation $f_S$, as demonstrated in Eq. \ref{eq4}.

\begin{equation}
\label{eq4}
f_S= \mathcal{M}_{S} \left [ ..., x_{t-1}(n) , x_{t}(n), x_{t+1}(n), ... \right ]     
\end{equation}

%

\textbf{Definition 3} (\textit{Multi-level Device Feature Fusion}). Multi-level device feature fusion refers to the use of multiple layers of device features to integrate the feature information of different layers through a reasonable fusion method, which can give full play to the strengths of each layer, make up for the shortcomings of each layer, and achieve a higher level of device performance. The outputs of different branches or different layers of the network are usually used as representation features of different layers and spliced using a reasonable method, as demonstrated in Eq. \ref{eq5}.

\begin{equation}
\label{eq5}
f_C = \mathcal{C}_{S}\left [ f_{F}, f_S\right ]       
\end{equation}

\begin{figure*}[h]
	\centering
	\includegraphics[width=\linewidth]{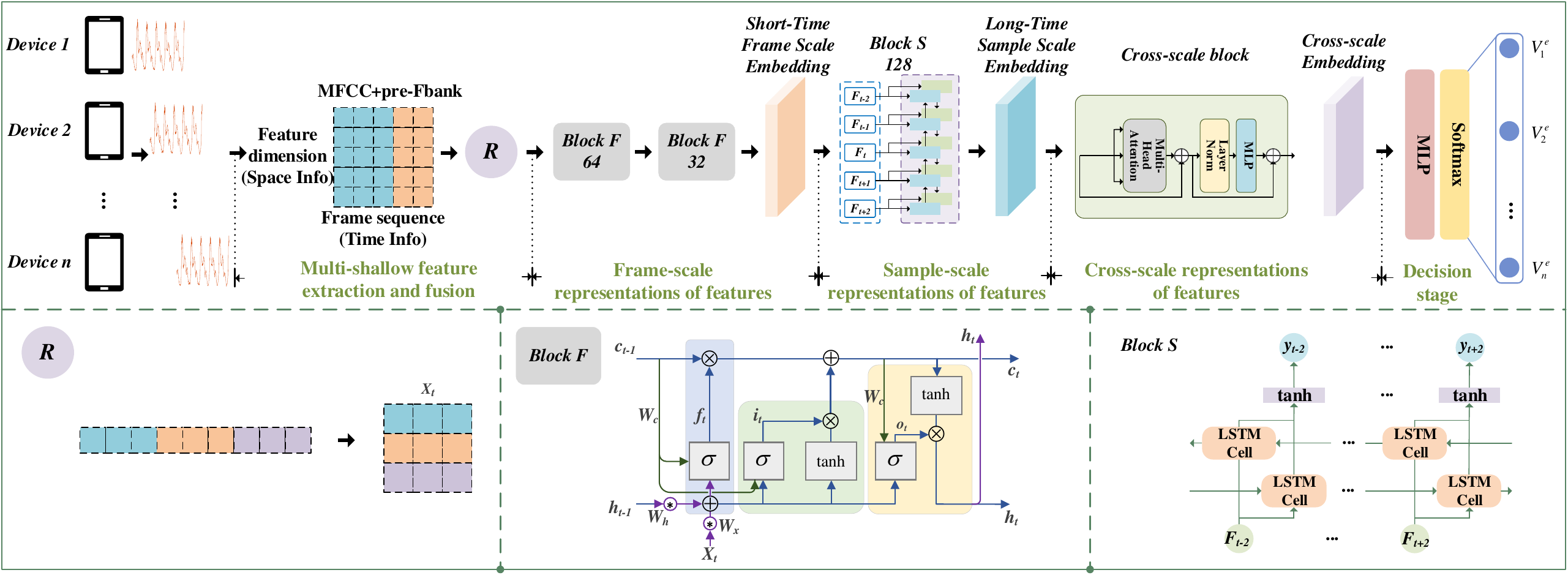}
	\caption{Overall framework of the proposed cross-scale and multi-level representation learning based mobile recording device recognition method.}
	\label{fig_2}
\end{figure*}

\section{Proposed Method} \label{4}

To facilitate the backend modeling in capturing more representative depth information from audio signals, this study employs a multi-level feature extraction approach in the frontend. This strategy optimally leverages information obtained at multiple processing stages in the frequency domain of audio signals. In contrast to utilizing solely the final frequency domain features, the multi-level frontend features provide a more comprehensive representation of audio signals.

Regarding the two-dimensional temporal features extracted in the frontend, this paper designs the backend network model from three perspectives: frame-level scale, sample-level scale, and global feature fusion. Initially, spatiotemporal feature enhancement fusion at the frame-level scale is achieved using ConvLSTM based on one-dimensional convolution. Following this, BiLSTM is employed to learn the representation of sample-level sequence information. Subsequently, the features undergo processing at the global scale through the Transformer-encoder, and the discriminative results are finally obtained through softmax. The overall framework of the proposed approach is depicted in Fig. \ref{fig_2}. This chapter will present the descriptions in the order of frame-level feature processing, sample-level feature processing, and global information processing.

\subsection{Timing Tandem Features Based on MFCC Multi-level Localized Frame-levels}

Analyzing audio signals in the frequency domain or through cepstral analysis is a process that gradually extracts information from lower to higher layers. In the initial stages, features capture relatively basic information, which is then refined through manual transformations to achieve more targeted feature extraction.

\begin{figure}[h]
\centering
\includegraphics[width=3.45in]{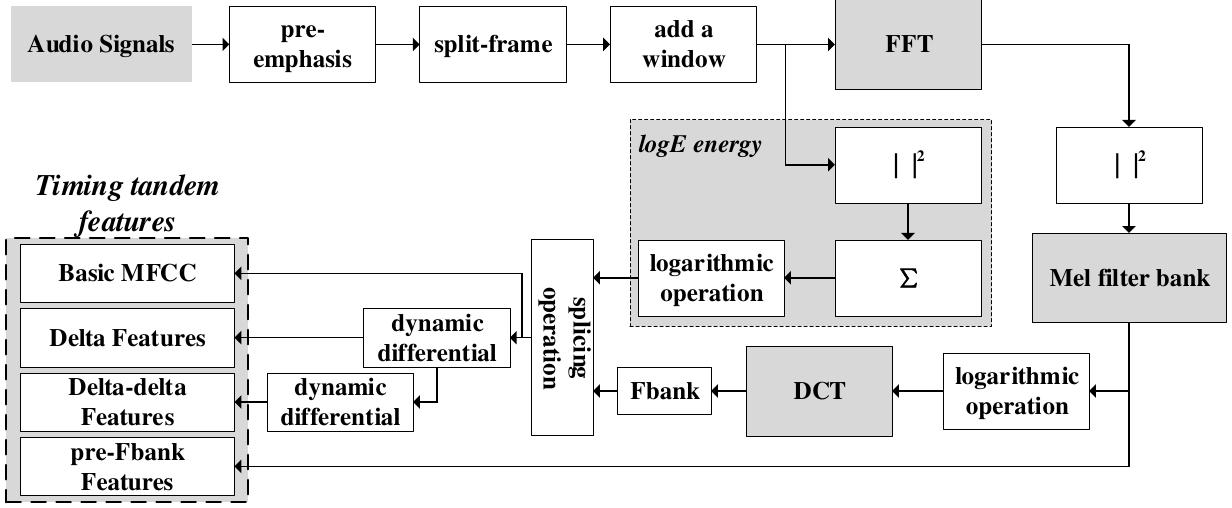}
\caption{Process of MFCC-based timing tandem feature extraction.}
\label{fig_3}
\end{figure}

MFCC, as a crucial feature of audio signals, emphasizes information carried by the low-frequency range, aligning with the frequency sensitivity of human auditory perception. Its extraction process involves filtering the frequency-domain information, obtained through Fast Fourier Transform (FFT), with a set of bandpass triangular filters. The Fbank feature coefficients are then derived from the power spectrum signal through a Mel filterbank. After nonlinear correction through logarithmic operations, the Fbank undergoes discrete cosine transform (DCT) to yield the fundamental MFCC feature. In the extraction process, LogE energy coefficients are also calculated based on frame energy, representing the logarithm of the sum of squares of all elements in a frame and measuring the energy magnitude of the speech frame. Delta features, representing the dynamic differences between frames and obtained through the temporal differentiation of the base MFCC, as well as the second-order differential coefficients, Delta-delta features, are additional dynamic change indicators between frames. They reflect the temporal status changes of the basic MFCC features and serve as two sets of auxiliary features for training the network model, contributing to better accuracy and generalization ability. Pre-Fbank features represent the characteristics before logarithmic operations and DCT, retaining relatively complete original information. The entire process is illustrated in Fig. \ref{fig_3}. To enhance reader comprehension, we present the following process as an algorithmic table in Algorithm \ref{algorithm1}.

\begin{figure}[h]
\centering
\includegraphics[width=3.45in]{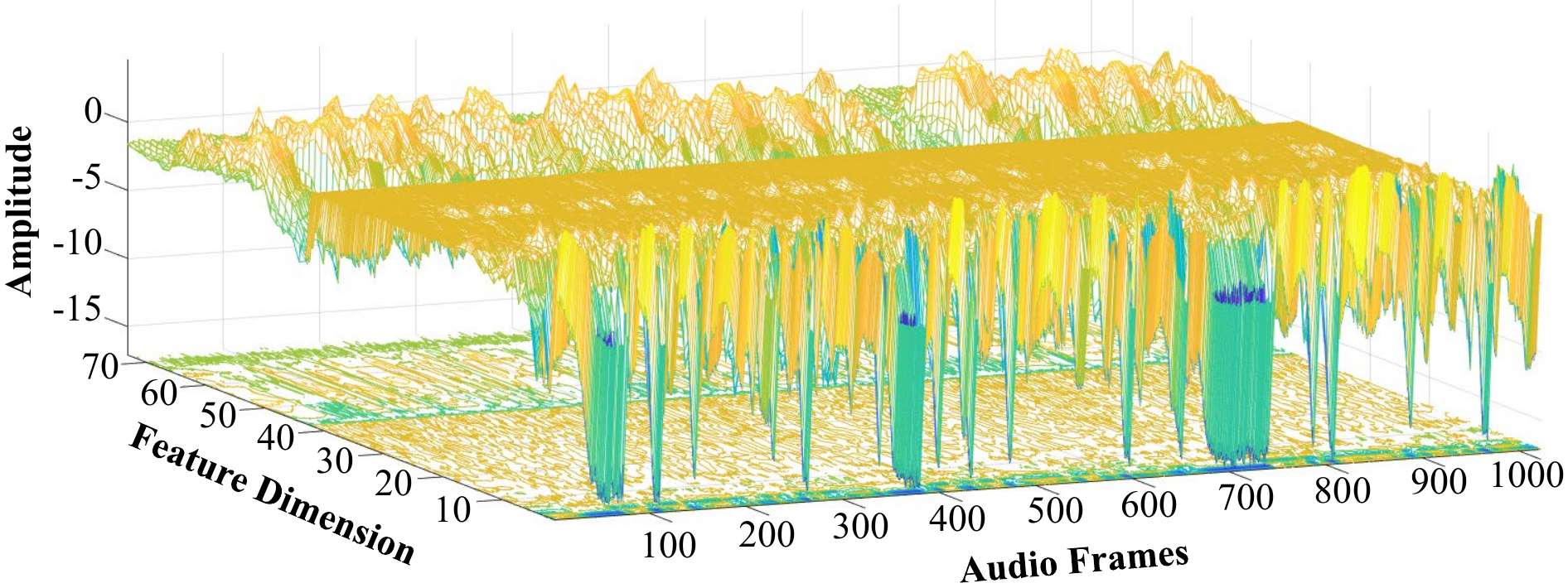}
\caption{Schematic of the dynamic range of the timing tandem feature.}
\label{fig_4}
\end{figure}

The application of multi-shallow MFCC features serves to enrich the information available for further refinement and enhancement within the deep learning network's backend. This paper illustrates the dynamic range of temporally concatenated features through Fig. \ref{fig_4}. Notably, the temporally concatenated features, derived from multi- shallow MFCC extraction, exhibit a broader dynamic range and offer richer information than the basic MFCC. These frontend features, when fed into the backend network model, facilitate the analysis of connections among different hierarchical information within individual sequences. Beyond examining the dynamic information contained in the first and second-order difference coefficients, the model can also learn to represent long-term information for pre-Fbank features and basic MFCC across extended sequences. This approach results in a comprehensive representation that ultimately improves discriminative accuracy.

\begin{algorithm}[htbp]
	\caption{Multi-shallow localized temporal tandem device feature extraction process.} \label{algorithm1}
    \LinesNumbered
    \KwIn{Speech data $x(n)$}
	
	\KwOut{Sample-level features}
    \For{$k=1:N$}{
        Pre-emphasis\
        \[y(n)=x(n)-\alpha \cdot x(n-1)\]\\
        Split-frame\
        \[y^{i}(n),i=0,1,2,...,f_{s}\]\\
        Add window functions\
        \[y_{1}^{i}(n)=w(n)\cdot y^{i}(n)\]\\
        Simultaneous computation of FFT and logarithmic energy\
        \[y_{3}^{i}(n)=log\sum_{i=1}^{f_{s}}\left | y_{1}^{i}(n)  \right |^{2}   \]\\
        Through the Mel Scale Filter Bank\
        \[y_{4}^{i}(n)=\left | y_{2}^{i}(n)  \right |^{2}\cdot W_{Mel}(n)\]\\
        DCT transform\
        \[y_{5}^{i}(n)=DCT\left \{ logy_{4}^{i}(n)  \right \}\]\\
        Stitching to get the final result\
        \[Features=\left [ y_{3}^{i}(n),y_{5}^{i}(n)   \right ]\]\\
    }
\end{algorithm} 

\subsection{Modeling Mobile Recording Device Recognition Based on Multi-Level Scales}\label{4.2}

In this section, a trainable network model is utilized to comprehensively capture features extracted at three levels: frame, sample, and global. The one-dimensional ConvLSTM captures both temporal and spatial information for frame-level spatiotemporal feature learning. The bidirectional structure of BiLSTM enhances the temporal learning for long-term sample-level features. The multi-head attention mechanism in the Transformer-encoder facilitates global interaction for deep feature processing at the final step.

\subsubsection{Spatiotemporal Fusion Augmentation of Short-term Frame-level Localized Features Based on 1D ConvLSTM}\label{4.2.1}

The frontend feature spectrogram, reshaped with the x-axis representing temporal information and the y-axis denoting the discrete frequency information of the audio input, exhibits a structure akin to image data, showcasing robust spatial characteristics. This design is particularly conducive to leveraging convolutional operations for the extraction of spatial features. ConvLSTM incorporates convolutional structures in both input-state and state-state transitions, facilitating the encoding of positional information from input data and hidden states. This integration effectively embeds spatial correlations into sequence data. The formula for ConvLSTM is presented below:
\begin{equation}
\label{eq6}
i_{t}=\sigma\left(W_{i x} * x_{t}+W_{i h} * h_{t-1}+W_{i c} \odot c_{t-1}+b_{i})\right.  
\end{equation}
\begin{equation}
\label{eq7}
f_{t}=\sigma\left(W_{f x} * x_{t}+W_{f h} * h_{t-1}+W_{f c} \odot c_{t-1}+b_{f})\right.  
\end{equation}
\begin{equation}
\label{eq8}
o_{t}=\sigma\left(W_{o x} * x_{t}+W_{o h} * h_{t-1}+W_{o c} \odot c_{t}+b_{o})\right.  
\end{equation}
\begin{equation}
\label{eq9}
g_{t}=tanh\left(W_{g x} * x_{t}+W_{g h} * h_{t-1}+b_{g})\right.  
\end{equation}
\begin{equation}
\label{eq10}
c_{t}=f_{t}\odot c_{t-1}+i_{t}\odot g_{t}       
\end{equation}
\begin{equation}
\label{eq11}
h_{t}=o_{t}\odot tanh(c_{i} )            
\end{equation}

In these equations, $x_{t}$, $c_{t}$, $h_{t}$, $i_{t}$, $f_{t}$,and $o_{t}$ are 2D tensors, where the first dimension represents temporal information at the sequence scale, and the second dimension represents spatial information at the feature scale.

\begin{figure}[h]
\centering
\includegraphics[width=3.5in]{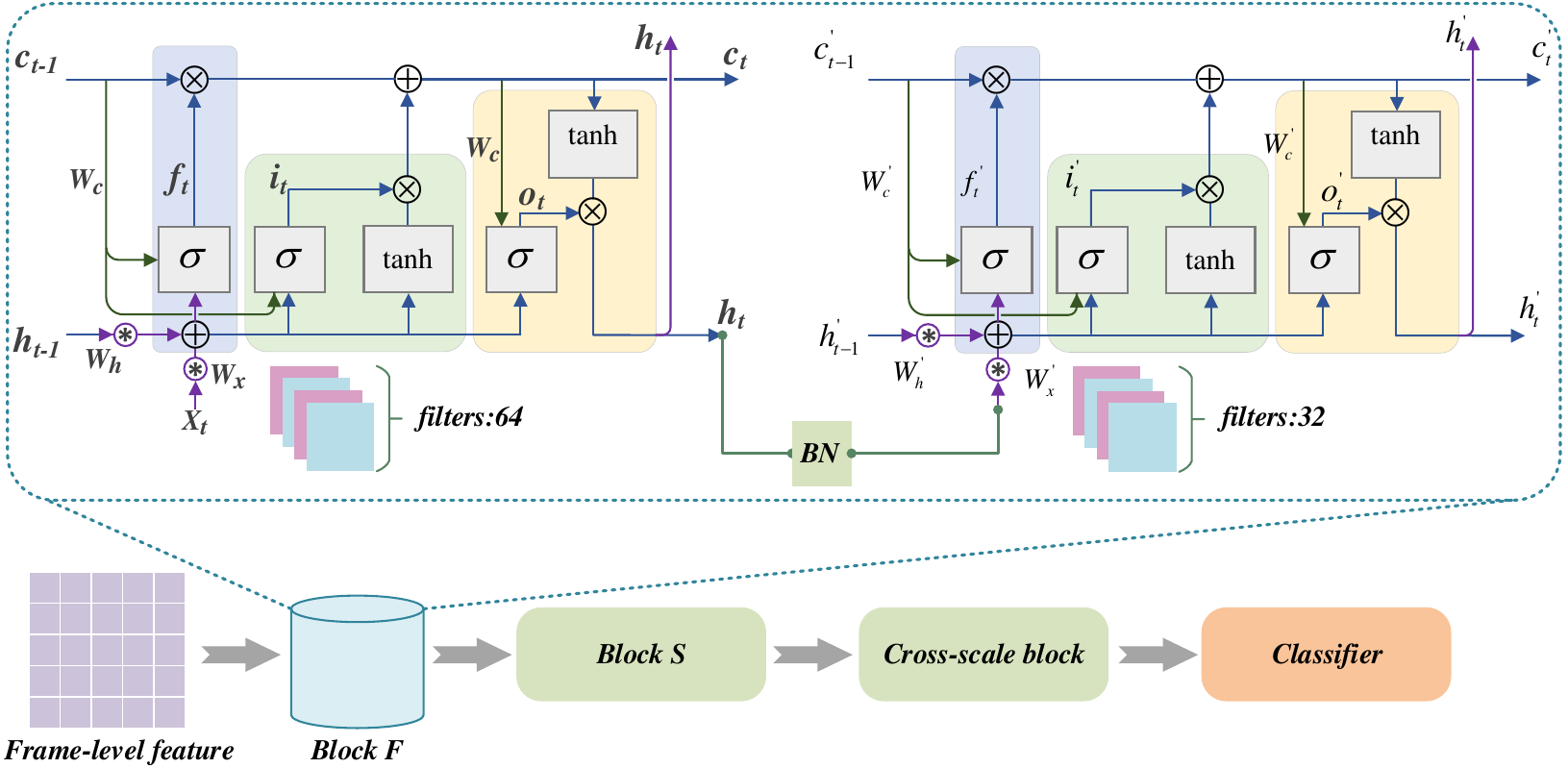}
\caption{Schematic diagram of the module for processing short-time frame-level features.}
\label{fig_5}
\end{figure}

This paper incorporates two ConvLSTM layers as the frame-level processing block. The first layer comprises 64 convolutional kernels, while the second layer consists of 32 convolutional kernels. Adopting a "larger kernel size in the upper layer, smaller in the lower layer" architecture facilitates reasonable information filtration, enhancing spatiotemporal fusion and in-depth exploration of original features at the frame level. Batch normalization is applied between the two modules to expedite model training. The internal structure of this segment is depicted in Fig. \ref{fig_5}.

\subsubsection{BiLSTM-based Representation Learning on Long-term Sample-level Feature Scales}\label{4.2.2}

After enhancing the features through one-dimensional ConvLSTM fusion, they are fed into a BiLSTM layer for long-term representation learning at the sample-level. The data, post feature fusion enhancement, maintains the format of two-dimensional temporal features. The sequence length remains constant, while the feature vector length is determined by the product of the length after one-dimensional convolution and the number of convolutional kernels.

\begin{figure}[h]
\centering
\includegraphics[width=3.5in]{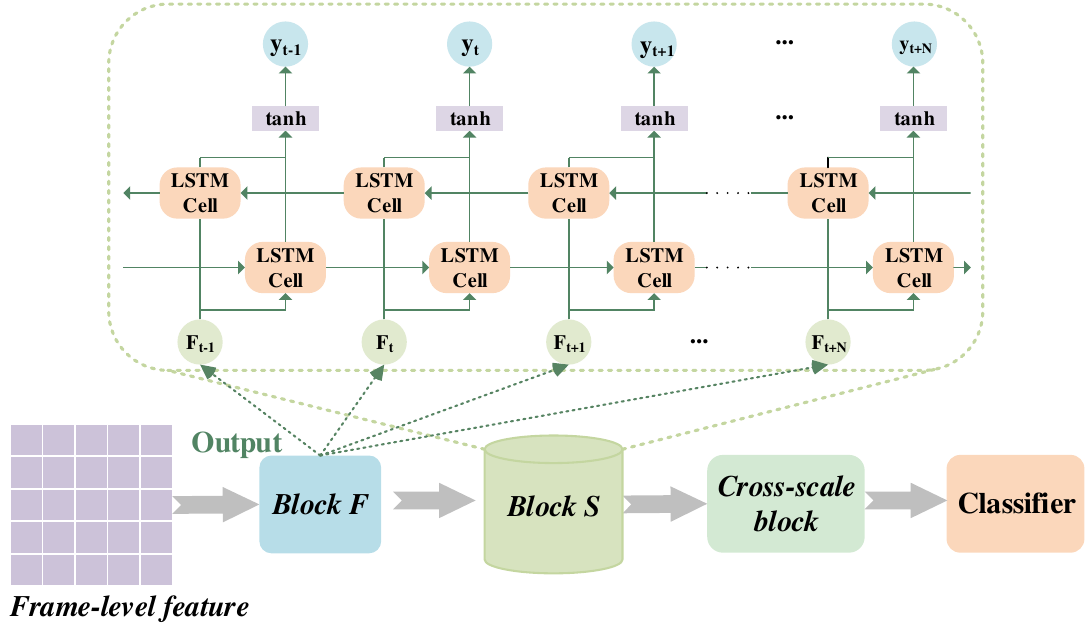}
\caption{Schematic diagram of the device characterization module at the processing sample-level.}
\label{fig_6}
\end{figure}

In this study, a BiLSTM module with 128 LSTM units is employed as the sample-level processing block. The BiLSTM layer consists of sequences of forward and backward LSTM units to extract short-term correlations and long-term dependencies in both directions. In each direction, the short-term correlations and long-term dependencies extracted from the feature data are updated to hidden units. The hidden units of the forward and backward sequences are concatenated into a sequence at the hidden gate, and the last hidden unit of each bidirectional sequence is linked to the subsequent Transformer-encoder. Fig. \ref{fig_6} illustrates the structure of the module for sample-level feature processing.

The BiLSTM layer is set to produce a sequence structure in its output, specifically outputting only the hidden state of the last time step. This configuration signifies that the ultimate neural output is a one-dimensional tensor of the output feature dimension.

\subsubsection{Transformer-encoder-based Learning of Short and Long-term Representations at Global Feature Scales}

Originally designed for handling word sequences in natural language processing, Transformer has successfully found applications in diverse domains, yielding impressive results. This success is attributed to Transformer's expansive global receptive field, enabling it to capture long-term contextual information in audio signals. Furthermore, the multi-head attention mechanism plays a crucial role in facilitating parallel processing of sequential data \cite{bib31}.

As indicated in Sections \ref{4.2.1} and \ref{4.2.2}, the model's frontend features undergo a Reshape operation, highlighting their spatial information structure, similar to the treatment of image data. Following the processing through the initial two modules, dedicated operations for frame-level and sample-level are conducted, accompanied by spatiotemporal fusion enhancement, ensuring efficient information extraction. However, the concatenation of these two modules predominantly emphasizes frame-level and sample-level information, underscoring the need for a global perspective on information interaction. The Transformer-encoder, with its remarkable multi-head self-attention mechanism, excels in aggregating global information. The multi-head attention mechanism, an innovation in self-attention, incorporates three trainable parameter matrices in each layer: the query matrix ($Q$), key matrix ($K$), and value matrix ($V$). The attention mechanism is mathematically represented as follows:
\begin{equation}
\label{eq12}
Attention(Q,K,V)=softmax(\frac{QK^{T} }{\sqrt{d_{k} } } )V            
\end{equation}

Initially, three parameter matrices $Q$, $K$, and $V$ are initialized based on the input features $X$. The dot product of the query matrix $Q$ and key matrix $K$ is computed to generate a similarity matrix. Subsequently, softmax normalization is applied to obtain similarity weights. These weights are then used to perform a weighted sum on the value matrix $V$, resulting in features with reinforced representation. Building on this foundation, the multi-head attention mechanism divides the features into distinct segments. Each segment undergoes independent self-attention computation. In comparison to a single-head attention mechanism, the multi-head variant can extract features at multiple hierarchical levels, implementing attention mechanisms for multiple pieces of information. The formula for the multi-head attention mechanism can be represented as:
\begin{equation}
\label{eq13}
head_{h}=Attention(QW_{h}^{Q},KW_{h}^{k},VW_{h}^{V}   )             
\end{equation}
\begin{equation}
\label{eq14}
MultiHead(Q,K,V)=\left [ head_{1},...,head_{H}  \right ]W^{O}               
\end{equation}

Here, $W^{O}$ represents the weight matrix computed through global training, utilized to reduce the dimensionality back to the target dimension.

For the same input, the multi-head attention mechanism defines multiple sets of distinct $W^{Q}$, $W^{K}$,and $W^{V}$. Each set calculates and generates different sets of $Q$, $K$, and $V$.The computation process of the attention mechanism is repeated several times to enable the model to extract information from different subspaces, ultimately facilitating the learning of distinct parameters.

Following the multi-head attention mechanism layer, the output is connected to the input features via a residual connection. Layer Normalization (LN) is applied, which, in contrast to conventional normalization, normalizes the input layer across feature dimensions. The subsequent formula illustrates the process of calculating layer normalization for all hidden units within the same layer:
\begin{equation}
\label{eq15}
\mu^{l}=\frac{1}{H} \sum_{i=1}^{H} a_{i}^{l}               
\end{equation}
\begin{equation}
\label{eq16}
\sigma ^{l}=\sqrt{\frac{1}{H}\sum_{i=1}^{H}(a_{i}^{l}-\mu ^{l}  )^{2}   }                 
\end{equation}

The concluding segment of the feedforward network constitutes the output layer of the Transformer-encoder, comprising a MultiLayer Perceptron (MLP).

\begin{figure}[!t]
\centering
\includegraphics[width=3.5in]{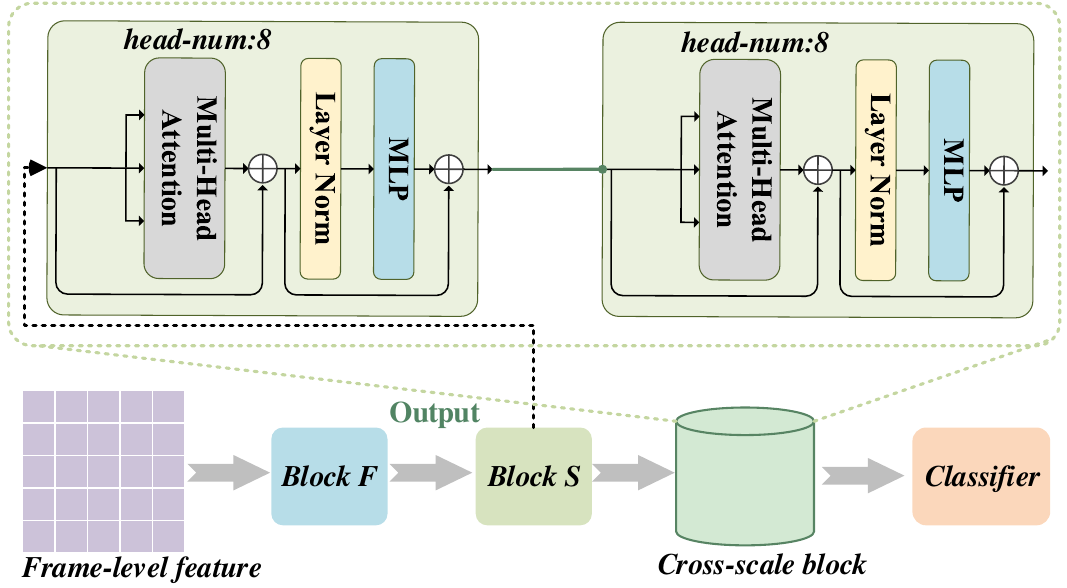}
\caption{Schematic diagram of the global information processing module.}
\label{fig_7}
\end{figure}

In this study, the Transformer-encoder functions as a global information interaction module (Cross-scale block), learning a global-scale representation of the discrete frequency information acquired from BiLSTM at the global feature scale. Notably, when utilizing the Transformer encoder, the initial LN layer is omitted. This decision is driven by two primary reasons: firstly, and most significantly, our aim is to fully retain the spatiotemporal information enhanced by the preceding modules; secondly, the LN layer typically addresses the issue of varying sequence lengths in sequence samples, whereas the features we extract are inherently frame-aligned. Fig. \ref{fig_7} outlines the structure of this module in the current research. To articulate this process more clearly, we have designed the following Algorithm \ref{algorithm2}.
\begin{algorithm}[htbp]
	\caption{Cross-scale Information Processing Module Algorithm Flow.} \label{algorithm2}
    \LinesNumbered
    \KwIn{Hidden layer state at the last time step of BiLSTM}
	
	\KwOut{Multi-level and multi-scale depth characterization}
    \For{$k=1:H$}{
        Initialize $Q,K,V$ for multi-segment features and calculate Attention\
        \[\left\{\begin{matrix}
 head_{h}=Attention(QW_{h}^{Q},KW_{h}^{k},VW_{h}^{V}   )\\MultiHead(Q,K,V)=\left [ head_{1},...,head_{H}  \right ]W^{O}

\end{matrix}\right.\]\\
       
    }
    Adding the residual structure through LN\
    \[h^{t}=f\left [ \frac{g}{\sigma ^{t} }\cdot (MultiHead^{t}-\mu ^{t})+b     \right ]\]\\
    Add window functions\
    \[y_{1}^{i}(n)=w(n)\cdot y^{i}(n)\]\\
    Through MLP\
    \[h_{1}^{t}=(w^{1}h^{t}+b^{1}   )\]\\
    Adding the residual structure through LN\
    \[Output=f\left [ \frac{g^{1} }{\sigma _{1}^{t} }\cdot (h_{1}^{t}-\mu_{1} ^{t})+b^{2}     \right ]\]\\ 
\end{algorithm} 

This study concatenates two 8-headed Transformer-encoders for cross-scale analysis, specifically focusing on learning representations of discrete frequency information from unrelated sequences. The input to the Transformer-encoder consists of the output of the last time step's hidden state from each BiLSTM. Following the analysis of long-term sample-level features, the temporal structure is eliminated, resulting in a one-dimensional feature vector composed of stacked frequency information from all sequences. The Transformer encoder processes this feature vector synchronously and in parallel. After positional encoding of the input feature vector, the multi-head attention mechanism is employed to explore global information within the features, effectively simulating short-term dependencies among discrete features.
\subsection{Training and optimization}

In the deep learning model experiments discussed in this paper, the recognition results of samples during the testing phase are determined by the model's final layer, computed using a Softmax layer. The Softmax layer maps the outputs of each neuron in the penultimate fully connected layer to the (0,1) interval. This process generates scores for each class in a multi-class classification task, calculating the probabilities of belonging to each class and, consequently, establishing the recognition result. For a total of $n$ numerically represented classes $S_{k}$, where $n$ denotes the number of classes and $k \in (0,n]$, the Softmax calculation formula is:
\begin{equation}
\label{eq17}
P(S_{i} )=\frac{e^{g_{i} } }{\sum_{k}^{n}e^{g_{k} } }                  
\end{equation}

Here, $i$ denotes a specific category within $k$, and $g$ represents the value associated with that category. The cumulative Softmax values for each category ultimately equal 1.

Simultaneously, we opted for the categorical crossentropy loss function, which is defined by the following formula:

\begin{equation}
\label{eq18}
Loss=-\sum_{i=1}^{outputsize}y_{i}\cdot log\hat{y_{i} }                     
\end{equation}

Cross-entropy loss measures the discrepancy between the model and the true labels by calculating relevant numerical values in the probability distribution. As evident from the formula above, $y_{i}$ has only two possible outcomes, 0 or 1. Therefore, the formula yields a result only when $y_{i}=1$. In simpler terms, categorical crossentropy exclusively focuses on the classification scenario where there is only one correct outcome.

Concerning the choice of optimizer, this study opts for the Adaptive Moment Estimation (Adam) optimizer, a form of adaptive learning rate optimizer based on the gradient descent algorithm. Adam combines the momentum algorithm with the introduction of two exponentially weighted values, enabling it to compute distinct learning rates for each parameter. This adaptability enhances the model's generalization capability while maintaining relatively low computational costs, contributing to the widespread use of Adam in training deep models.

Algorithm \ref{algorithm3} provides an overview of the training process in this study, offering a comprehensive view of the steps involved in model construction. The algorithm illustrates forward propagation, loss computation, momentum parameter calculation, and multiple historical parameter updates. The practicality of using the Adam optimizer for optimization adjustments is demonstrated, emphasizing its effectiveness in the training process.
\begin{algorithm}[htbp]
	\caption{Algorithm for the overall model-building process.} \label{algorithm3}
    \LinesNumbered
    \KwIn{Input features $x(n)$}
    Initialization weights $W_{all}$\\
    \For{$k=1:Epoch$}{
        Forward propagation\
        \[Softmax(MLP(Encoder(Block S(Block F(x))))\]\\
        Calculate the loss\
        \[Loss=-\sum_{i=1}^{outputsize}y_{i}\cdot log\hat{y_{i} }\]\\
        Calculate  momentum parameters\
        \[\left\{\begin{matrix}
  m_{t}=\beta _{1}m_{t-1}+(1-\beta _{1} )g_{t}\\
  v_{t}=\beta _{2}v_{t-1}+(1-\beta _{2} )g_{t}^{2}
\end{matrix}\right.\]\\
        Parameter correction\
        \[\left\{\begin{matrix}
  \hat{m} _{t}=\frac{m_{t} }{1-\beta _{1}^{t} } \\
  \hat{v} _{t}=\frac{v_{t} }{1-\beta _{2}^{t} } 
\end{matrix}\right.\]\\
        Update\
        \[W_{all}^{t+1}=W_{all}^{t}-\frac{\eta }{\sqrt{\hat{v_{t} } }+\varepsilon  }\hat{m_{t}}\]\\
    }
\end{algorithm}

\section{Experiment Results and Analysis}  \label{Exp}

To comprehensively assess the proposed method, this paper conducts experiments to address three fundamental research questions:

\textbf{\textit{RQ1:}} Does the proposed model outperform baseline systems in terms of effectiveness?

\textbf{\textit{RQ2:}} What roles do each blocks play within the proposed model? Are these modules compatible in terms of functionality?

\textbf{\textit{RQ3:}} Can the proposed model be effectively transferred between two different datasets?

To address these questions, three sets of experiments are designed in Sections \ref{5.6}---\ref{5.8}: (1) Comparative experiments with baseline algorithms, (2) Single-model and double-model ablation experiments exploring the roles of each module, and (3) Transferability performance experiments for the proposed model. To provide a clear experimental context, Sections \ref{5.1}---\ref{5.5} outline considerations related to experiments, including hardware and software devices, datasets, evaluation metrics, benchmarks, and hyperparameter settings.

\subsection{Experimental Platform Setup} \label{5.1}
The experiments are primarily conducted on a desktop computer with a Windows 10 operating system. This computer is equipped with a TITAN RTX GPU, an Xeon(R) Gold 5218 CPU, and 32 GB of RAM. The software used in the experiments mainly includes MATLAB 2015b, PyCharm, and Anaconda, along with toolkits such as Voicebox, MSR Identity Toolkit, and TensorFlow.

\subsection{Datasets} \label{5.2}
This paper conducted experiments using two databases: the CCNU\_Mobile dataset and the MOBIPHONE dataset.

The CCNU\_Mobile dataset \cite{bib32} comprises audio data recorded by 45 devices of different models. The recording devices come from 9 different brands, including Apple, Huawei, Honor, Nubia, Oppo, Vivo, Xiaomi, Samsung, and ZTE. Detailed information about the device models is available in Table \ref{tab:table2}.

The recording corpus of this dataset is sourced from the Texas Instruments/Massachusetts Institute of Technology (TIMIT) dataset. In the recording process of the CCNU\_Mobile dataset, all training data from the TIMIT dataset was concatenated into a single long audio file with a duration of approximately 110 minutes. Subsequently, this long audio file was played in a quiet, dedicated recording studio environment, and recordings were made simultaneously using 45 devices. After recording, the long audio file was segmented, resulting in 642 audio samples recorded by each device. Each recording sample has a duration of approximately 10 seconds, stored as single-channel audio files in wav format, with a sampling rate of 32000 Hz and a bit rate of 512 kbps.

In the experiments conducted in this paper, 514 audio samples from each category of recording devices were selected as the training set, while the remaining 128 audio samples were used as the test set. The training set and test set constitute 80\% and 20\% of the total samples, respectively, with the validation set being 20\% of the training set.
\begin{table}[h]
\caption{CCNU\_Mobile dataset device models. \label{tab:table2}}
\centering
\begin{tabular}{cc}
\hline
Brands  & Models                                                                                                                 \\ \hline
APPLE   & \begin{tabular}[c]{@{}c@{}}iphone6(4), iphone6s(3), iPhone SE, \\ ipad7, iphone7p, iphoneX, air2(2), air1\end{tabular} \\
HUAWEI  & Nova, Nova2s, Nova3e, P10, P20, TAG-AL00                                                                               \\
HONOR   & honor7x, honor8(3), honorV8, honor9, honor10                                                                           \\
XIAOMI  & \begin{tabular}[c]{@{}c@{}}mi2s, note3, mi5, mi8, mi8se(2),\\ mix2, redmiNote4x, redmi3S\end{tabular}\\
VIVO    & y11t,x3f,x7                                                                                                            \\
ZTE     & c880a,g719c                                                                                                            \\
SAMSUNG & sphd710,s8                                                                                                             \\
OPPO    & r9s                                                                                                                    \\
NUBIA   & z11                                                                                                                    \\ \hline
\end{tabular}
\end{table}

In Section \ref{5.8}, when investigating the transfer performance of the proposed model, this paper utilized the MOBIPHONE dataset \cite{bib33}. This dataset, recorded by Kotropoulos et al. in 2014, is a publicly available speech dataset suitable for testing the model's transferability. The dataset was created using the TIMIT corpus with 21 different models of devices in a quiet environment, featuring speech content from 12 male and 12 female speakers. Audio samples are stored in wav format with a sampling rate of 16 kHz and a bit depth of 16 bits.

This paper selected 120 speech samples from each category of recording devices as the training set and 20 speech samples as the test set, with an 8:2 ratio between the training set and the validation set, as shown in Table \ref{tab:table3}.

\begin{table}[h]
\caption{MOBIPHONE dataset device models. \label{tab:table3}}
\centering
\begin{tabular}{cc}
\hline
Brands   & Models                                                                                                                                    \\ \hline
HTC      & desire c, sensation xe                                                                                                                    \\
LG       & GS290,L3,Optimus L5,Optimus L9                                                                                                            \\
Nokia    & C5,N70                                                                                                                                    \\
Sony     & Ericsson c902,Ericsson c510i                                                                                                              \\
APPLE    & iPhone5                                                                                                                                   \\
Samsung  & \begin{tabular}[c]{@{}c@{}}E2121B,E2600,GT-18190mini,GT-N7100(Galaxy Note2),\\ Galaxy GT19100 s2,Galaxy Nexus S,e1230,s5830i\end{tabular} \\
Vodafone & Joy 845                                                                                                                                   \\ \hline
\end{tabular}
\end{table}

\subsection{Evaluation Metrics} \label{5.3}

In classification problems, prediction results typically fall into four categories: a) True Positive (TP); b) False Positive (FP); c) True Negative (TN); d) False Negative (FN). 

For a multi-class model, using the first class as an example (denoted by subscript 1), TP represents the number of samples that actually belong to the first class and are correctly predicted as such. FP indicates the number of samples that do not belong to the first class but are incorrectly predicted as belonging to it. TN denotes the number of samples that are not part of the first class and are correctly identified as such. FN refers to the number of samples that actually belong to the first class but are incorrectly predicted as not belonging to it.

To facilitate a visual assessment of the model's recognition outcomes for each category, this paper utilizes the overall classification accuracy (ACC) as the primary performance metric, defined as:

\begin{equation}
\label{eq19}
ACC=\frac{S_{cr} }{S_{t} }\times 100\% 
\end{equation}
where $S_{t}$ denotes the total number of samples participating in the test, and $S_{cr}$ represents the count of samples correctly identified.

Imultaneously, this paper also opted for additional pertinent evaluation metrics for multi-class problems to gauge the model's performance (taking class 1 as an example), outlined as follows:

\textbf{Recall:}
\begin{equation}
\label{eq20}
Recall(c1)=\frac{TP_{1} }{TP_{1}+FN_{1}   }  
\end{equation}

This metric gauges the ratio of correct predictions among all samples of a specific type, emphasizing completeness, even in cases where some results may be incorrect.

\textbf{Precision:}
\begin{equation}
\label{eq21}
Precision(c1)=\frac{TP_{1} }{TP_{1}+FP_{1}  }   
\end{equation}

Precision is computed as the ratio of correct predictions among all samples predicted to belong to a specific category, prioritizing precision even if some predicted results are overlooked.

\textbf{F-Score:}
\begin{equation}
\label{eq22}
F_{\beta}(\operatorname{c} 1)\!=\!\left(1\!+\!\beta^{2}\right) \!\times\! \frac{\operatorname{Precision}(\operatorname{c} 1) \times \operatorname{Recall}(\operatorname{c} 1)}{\beta^{2} \!\times\! \operatorname{Precision}(\operatorname{c} 1)\!+\!\operatorname{Recall}(\operatorname{c} 1)}   
\end{equation}

$F_{\beta}-Score$ is a weighted harmonic metric that takes into account both recall and precision. The higher the value of this metric, the better the model's performance. Here, $\beta$ represents the harmonic weight assigned to recall and precision, indicating the varying degrees of importance attributed to each. In this experiment, we opted for the F1-score calculated with the weight specified by $\beta=1$.

\subsection{Baselines}\label{5.4}

To evaluate the overall performance of the proposed method, this section compares it with four traditional machine learning methods (MFCC+SVM \cite{bib4}, GSV+SVM\cite{bib11}, BED+SVM\cite{bib6} and i-vector+SVM\cite{Rao2013}) and three representative deep learning methods (GSV+CNN\cite{bib34} and Attention Multi-feature Fusion[\cite{bib32}, Multi-Attention\cite{Zeng2023}). A brief overview of the characteristics and parameter settings for the seveb benchmark methods is outlined below:

MFCC + SVM\cite{bib4}: This method utilizes MFCC, a widely adopted feature in audio recognition, in conjunction with SVM as the classification model. In this benchmark, 650 frames of MFCC vectors are extracted for each sample from the CCNU\_Mobile dataset. Each MFCC feature vector includes 13 dimensions of Mel cepstral coefficients, 13 dimensions of first-order differential signals, and 13 dimensions of second-order differential signals, resulting in an MFCC feature size of (640, 39).

GSV + SVM\cite{bib11}: GSV features are derived by concatenating mean vectors of each Gaussian mixture component in a GMM, transforming input feature spectral information into a probability density representation of audio segments. For this benchmark, GMM is trained using 39-dimensional MFCC with 64 Gaussian mixture components, resulting in a feature size of (39, 64) for the feature matrix.

BED + SVM\cite{bib6}:This method, based on spectral feature extraction, calculates the baseband energy difference to intuitively describe device source differences while effectively reducing computational overhead. In the BED feature extraction process, the number of Fourier spectrum sampling points is set to 256, and after differential operation, a baseband energy differential feature with dimensions (1, 127) is obtained.

i-vector + SVM\cite{Rao2013}:The i-vector reduces the dimensionality by obtaining the speech  feature vector of the high-dimensional target device source, projecting it in the subspace, and using factor analysis to eliminate the factors that put redundancy to obtain the low-dimensional feature vector.

GSV + CNN\cite{bib34}: Combining the representative GSV feature with CNN, this method represents deep learning approaches. The data format of GSV features is a two-dimensional matrix, processed using a CNN with a 3×3 kernel size. Batch normalization layers are applied for feature map processing between convolutional layers.

Multi-feature Fusion\cite{bib32}: This approach incorporates three input features – MFCC, GSV, and I-vector features. It utilizes CNN/Deep Neural Network (DNN) to extract bottleneck features from MFCC and GSV, followed by attention-based deep and shallow feature fusion with I-vector features.

Multi-Attention\cite{Zeng2023}:This approach develops a dual-branch network combining Residual Dense Time Convolutional Networks (RD-TCNs) and Convolutional Neural Networks (CNNs). It uses the spatial probability distribution features of audio signals as input for the CNN branch to facilitate spatial representation learning. Moreover, it incorporates three attention mechanisms—temporal, spatial, and branch—to capture spatiotemporal weights, thereby enabling effective deep feature fusion.

\subsection{Experimental Settings}\label{5.5}

When extracting temporal concatenated features based on MFCC at multiple hierarchical levels, the frame number is set to 128, indicating that each audio sample undergoes the extraction of a two-dimensional temporal feature with a sequence length of 128. A single concatenated feature vector is 73-dimensional, encompassing 12 dimensions of fundamental MFCC, 1 dimension of logE energy coefficient, 13 dimensions of first-order dynamic difference coefficient Delta, 13 dimensions of second-order dynamic difference coefficient Delta-delta, and 34 dimensions of pre-Fbank features.

In constructing the recognition model, the spatiotemporal enhancement at the frame level consists of two layers of one-dimensional ConvLSTM, where the number of convolutional kernels is set to 64 and 32. The size of the one-dimensional convolutional kernel is uniformly set to 3, with strides set to 3 and 2, respectively. Batch normalization layers are introduced after each layer of one-dimensional ConvLSTM to ensure the stability of the model's learning distribution. The resulting three-dimensional tensor with the structure (time sequence, feature length, number of convolutional kernels) is reshaped into a two-dimensional tensor with the structure (time sequence, feature length × number of convolutional kernels).

In the section dedicated to long-term representation learning of sample-level features, the input feature has the structure of a two-dimensional temporal feature (time sequence, feature length × number of convolutional kernels). The number of BiLSTM units is configured as 128, resulting in an output size of bidirectional 2×128=256. For short-term representation learning at the feature scale, the input feature is a one-dimensional vector with a length of 256. The number of Transformer-encoder is set to 2, with the matrix dimension of the multi-head attention mechanism set to 64, the number of heads set to 8, the MLP units set to 128, and the output forming a one-dimensional tensor composed of 128 neuron outputs.

The model optimizer uses Adam, the loss function is cross-entropy, and the initial learning rate is set to 0.0001, decreasing by a factor of 1/10 every 30 epochs. The distribution of the training set, validation set, and test set is allocated in a ratio of 64\%:16\%:20\%. The batch size during training is set to 64, and the training epochs are set to 100. The key parameters are summarized in the Table \ref{tab:table4}.
\begin{table}[h]
\caption{List of experiment-related parameter settings. \label{tab:table4}}
\centering
\begin{tabular}{cc}
\hline
Block                 & Parameter                                                                                                                                         \\ \hline
ConvLSTM              & \begin{tabular}[c]{@{}c@{}}$\left [ 3\times 3\ \ \ stride=3 \right ]\times 64$\\ $\left [ 3\times 3\ \ \ stride=2 \right ]\times 32$\end{tabular} \\ \cline{2-2} 
BiLSTM                & Units:128                                                                                                                                         \\ \cline{2-2} 
Transformer\_encoder  & \begin{tabular}[c]{@{}c@{}}Encoder\_num:2,\\ Head\_size:64,\\ Head\_num:8\end{tabular}                                                            \\ \cline{2-2} 
MLP                   & Units:128                                                                                                                                         \\ \cline{2-2} 
Other hyperparameters & \begin{tabular}[c]{@{}c@{}}Lr:1e-4,\\ Epoch:100\\ Batch\_size:64\end{tabular}                                                                     \\ \hline
\end{tabular}
\end{table}

\subsection{Comparison Experiments with the Baseline Algorithmic Model (RQ1)} \label{5.6}

To validate the sophistication of the proposed method, this section systematically compares it with the benchmark methods introduced in Section \ref{5.4}. These comparative experiments involve different configurations of feature data and modeling algorithms. Table \ref{tab:table5} outlines the features extracted by various methods at the front end and the classification models at the back end. The evaluation employs the metrics described in Section \ref{5.3}, and the forward time for one test sample is considered in the assessment.

\begin{table*}[h]
\caption{Comparison experiments with the baseline method. \label{tab:table5}}
\centering
\begin{tabular*}{0.85\linewidth}{ccccccc}
\hline
 Methods                       &  Features                       &  Models                      &  Accuracy             &  Precision             &  Recall             &  F1              \\ \hline
 MFCC + SVM\cite{bib4}             &  MFCC                           &  SVM                         &  86.8\%          &  86.9\%          &  86.8\%          &  86.9\%          \\
 GSV + SVM\cite{bib11}             &  GSV                            &  SVM                         &  97.6\%          &  97.8\%          &  97.6\%          &  97.5\%          \\
 BED + SVM\cite{bib6}              &  BED                            &  SVM                         &  93.4\%          &  93.7\%          &  93.4\%          &  93.3\%          \\
 i-vector+SVM\cite{Rao2013}                  &  i-vector                       &  SVM                         &  64.6\%                & 65.9\%                 & 64.6\%                 & 64.7\%                 \\
 GSV + CNN\cite{bib34}             &  GSV                            &  CNN                         &  92.9\%          &  93.6\%          &  92.9\%          &  92.8\%          \\
 Multi-feature Fusion\cite{bib32}  &  Three features                 &  CNN \& DNN                  &  97.2\%          &  97.4\%          &  97.2\%          &  97.1\%          \\
 Multi-Attention\cite{Zeng2023}               &  Spatio-temporal features       &  RD-TCN                      &  97.6\%          &  97.7\%          &  97.4\%          &  97.4\%          \\
 \textbf{Method of this paper} &  \textbf{Timing tandem feature} &  \textbf{BiLSTM-Transformer} &  \textbf{99.6\%} &  \textbf{99.6\%} &  \textbf{99.6\%} &  \textbf{99.6\%} \\ \hline
\end{tabular*}
\end{table*}

The proposed method in this paper achieved a recognition accuracy of 99.6\%, showcasing an improvement of 2.0\% to 16.5\% compared to the baseline methods. In other words, among the 5760 test samples for 45 classes, only 23 samples were misclassified. 

Table \ref{tab:table5} clearly indicates that models utilizing deep learning techniques generally outperform traditional machine learning models. Notably, among the three deep learning-based models listed, the proposed method in this paper attained the highest accuracy. The superiority of the proposed method is credited to the multi-level cross-scale processing of the original input data, resulting in more representative embeddings. Furthermore, the multi-head self-attention mechanism can effectively capture relevant information in the time series, exhibiting outstanding performance in recognizing the inherent temporal features of digital audio signals.

In practical applications, in addition to the evaluation metrics discussed in Section\ref{5.3}, it is essential to consider the computational costs of the proposed method. This section examines both time complexity and space complexity of the proposed and baseline methods. For time complexity, we measure the forward propagation time of an audio sample on the platform described in Section\ref{5.1}. For space complexity, we compare the number of model parameters for each method, with the results presented in Table\ref{tab:table6}.

\begin{table*}[h]
\caption{Experimental results comparing the proposed method and benchmark methods in terms of time complexity and space complexity. \label{tab:table6}}
\centering
\begin{tabular*}{0.85\linewidth}{ccccc}
\hline
 Methods                       &  Features                       &  Models                      &  \begin{tabular}[c]{@{}c@{}}Forward \\ sample time /ms\end{tabular} &  \begin{tabular}[c]{@{}c@{}}Model \\ parameter number /million\end{tabular} \\ \hline
 MFCC + SVM\cite{bib4}             &  MFCC                           &  SVM                         &  13.2                                                               &  0.102                                                                      \\
 GSV + SVM\cite{bib11}             &  GSV                            &  SVM                         &  30.6                                                               &  0.005                                                                      \\
 BED + SVM\cite{bib6}              &  BED                            &  SVM                         &  9.3                                                                &  0.019                                                                      \\
 i-vector+SVM\cite{Rao2013}                  &  i-vector                       &  SVM                         &  18.2                                                               &  0.022                                                                      \\
 GSV + CNN\cite{bib34}             &  GSV                            &  CNN                         &  2.0                                                                &  0.348                                                                      \\
 Multi-feature Fusion\cite{bib32}  &  Three features                 &  CNN \& DNN                  &  2.0                                                                &  2.361                                                                      \\
 Multi-Attention\cite{Zeng2023}               &  Spatio-temporal features       &  RD-TCN                      &  0.9                                                                &  40.560                                                                     \\
 Method of this paper &  Timing tandem featur &  BiLSTM-Transformer &  5.2                                                       &  0.626                                                             \\ \hline
\end{tabular*}
\end{table*}

Table \ref{tab:table6} clearly shows that the forward processing time for traditional machine learning methods is generally higher than for deep learning methods, due to the GPU acceleration used in the latter. However, traditional machine learning methods typically have fewer model parameters than deep learning methods. While the forward processing time for our proposed method is not the shortest, it remains within the same order of magnitude as models with a similar number of parameters, all while maintaining high accuracy. Although the Multi-Attention method achieves shorter forward processing times, it has approximately 15 times more parameters than our method. Overall, our proposed method demonstrates significant potential and practical value.

\subsection{Ablation Experiments Exploring the Role of Modules within a Model (RQ2)} \label{5.7}

The backend network of the proposed method comprises three network modules. To comprehensively assess each module's contribution to the model, two sets of ablation experiments are designed in this section. The first set (Group I) involves retaining only one network block in the backend and comparing it with the complete model. The second set (Group II) explores the impact of each pair of network blocks on overall performance. The design of these two ablation experiment sets is detailed below:

\textbf{Group I (Single-Module Ablation Experiment):} In this experimental set, each of the ConvLSTM, BiLSTM, and Transformer-encoder blocks is employed separately as the feature processing component of the backend network, as outlined in Table \ref{tab:table6} (Group 1-Group 3). This set of experiments allows for an intuitive analysis of the distinct roles played by each block in extracting temporal features.

\textbf{Group II (Double-Module Ablation Experiment):} In this experiment set, one block is removed from the backend network, retaining two blocks, as indicated in Table \ref{tab:table7} (Group 5-Group 7). The objective of these experiments is to analyze the overall impact of each block on the system level.

In both experiment sets, the input feature data remains consistent with Section \ref{5.5}, taking the form of two-dimensional temporal concatenated features with dimensions of (128, 73). In the ablation experiments, given the absence of certain blocks, dimension inconsistencies between layers may arise. To ensure efficient feature data transfer between layers with dissimilar input and output dimensions, Reshape or Flatten operations are employed to modify the dimension of feature data. Additionally, a fully connected layer facilitates the transition when connecting to the Softmax layer in the classification end.

\subsubsection{Single-Module Ablation Experiment Data and Analysis}\label{5.7.1}

\begin{figure}[!t]
\centering
\includegraphics[width=3.5in]{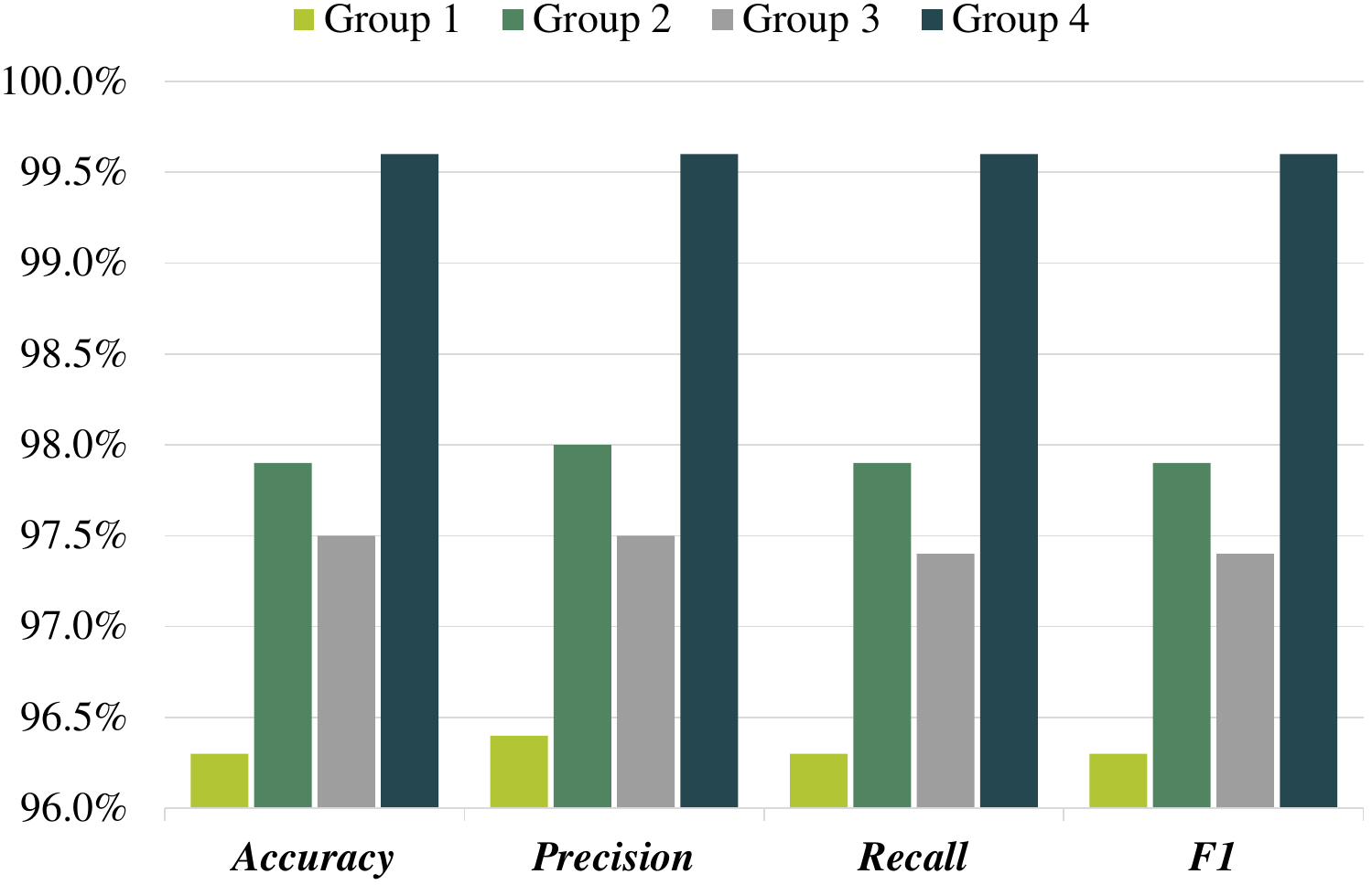}
\caption{Histogram comparing the data of single-module ablation experiments.}
\label{fig_8}
\end{figure}

Table \ref{tab:table7} presents the data resulting from the single-module ablation experiments, with the presence of each module denoted by "$\surd$". To distinctly depict the performance of each block under each metric and facilitate a more in-depth analysis and effective comparison, a histogram was generated to illustrate the experimental results, as depicted in Fig. \ref{fig_8}.

\begin{table*}[!t]
\caption{Single-module ablation experimental data. \label{tab:table7}}
\centering
\begin{tabular*}{0.645\linewidth}{cccc|cccc}
\hline
\multirow{2}{*}{} & \multirow{2}{*}{ConvLSTM} & \multirow{2}{*}{BiLSTM} & \multirow{2}{*}{Transformer} & \multicolumn{4}{c}{Evaluation metrics} \\ \cline{5-8} 
                  &                           &                         &                              & Accuracy      & Precision     & Recall     & F1      \\ \hline
Group1            &$\surd$                       &                         &                              & 96.3\%   & 96.4\%  & 96.3\%  & 96.3\%  \\
Group2            &                           & $\surd$                     &                              & 97.9\%   & 98.0\%  & 97.9\%  & 97.9\%  \\
Group3            &                           &                         & $\surd$                          & 97.5\%   & 97.5\%  & 97.4\%  & 97.4\%  \\
Group4            & $\surd$                       & $\surd$                     & $\surd$                          & 99.6\%   & 99.6\%  & 99.6\%  & 99.6\%  \\ \hline
\end{tabular*}
\end{table*}

From Table \ref{tab:table7} and Fig. \ref{fig_8}, it is evident that all three blocks can represent audio features to varying extents. Notably, optimal recognition results are attained through the feature processing of BiLSTM. BiLSTM excels at extracting sample-level features, while the Transformer module specializes in capturing global features, leveraging its global perspective for enhanced audio representation. On the contrary, ConvLSTM focuses on a more local perspective, extracting features at the frame-level, and its effectiveness in the single-module case falls behind the former two.

\subsubsection{Double-Module Ablation Experimental Data and Analysis}

Data for the double-module ablation experiment is presented in Table \ref{tab:table7}, where the symbol "$\surd$" retains the same meaning as explained in Section \ref{5.7.1}. The corresponding histogram is depicted in Fig.10.
\begin{table*}[!t]
\caption{Double-module ablation experimental data. \label{tab:table8}}
\centering
\begin{tabular*}{0.642\linewidth}{cccc|cccc}
\hline
\multirow{2}{*}{} & \multirow{2}{*}{ConvLSTM} & \multirow{2}{*}{BiLSTM} & \multirow{2}{*}{Transformer} & \multicolumn{4}{c}{Evaluation metrics} \\ \cline{5-8} 
                  &                           &                         &                           & Accuracy      & Precision     & Recall     & F1      \\ \hline
Group5            &                           & $\surd$                 & $\surd$                      & 97.8\%   & 97.8\%  & 97.8\%  & 97.7\%  \\
Group6            & $\surd$                   &                         & $\surd$                      & 60.4\%   & 62.5\%  & 60.4\%  & 59.5\%  \\
Group7            & $\surd$                   & $\surd$                 &                              & 98.9\%   & 98.8\%  & 98.8\%  & 98.8\%  \\ \hline
\end{tabular*}
\end{table*}
\begin{figure}[!t]
\centering
\includegraphics[width=3.5in]{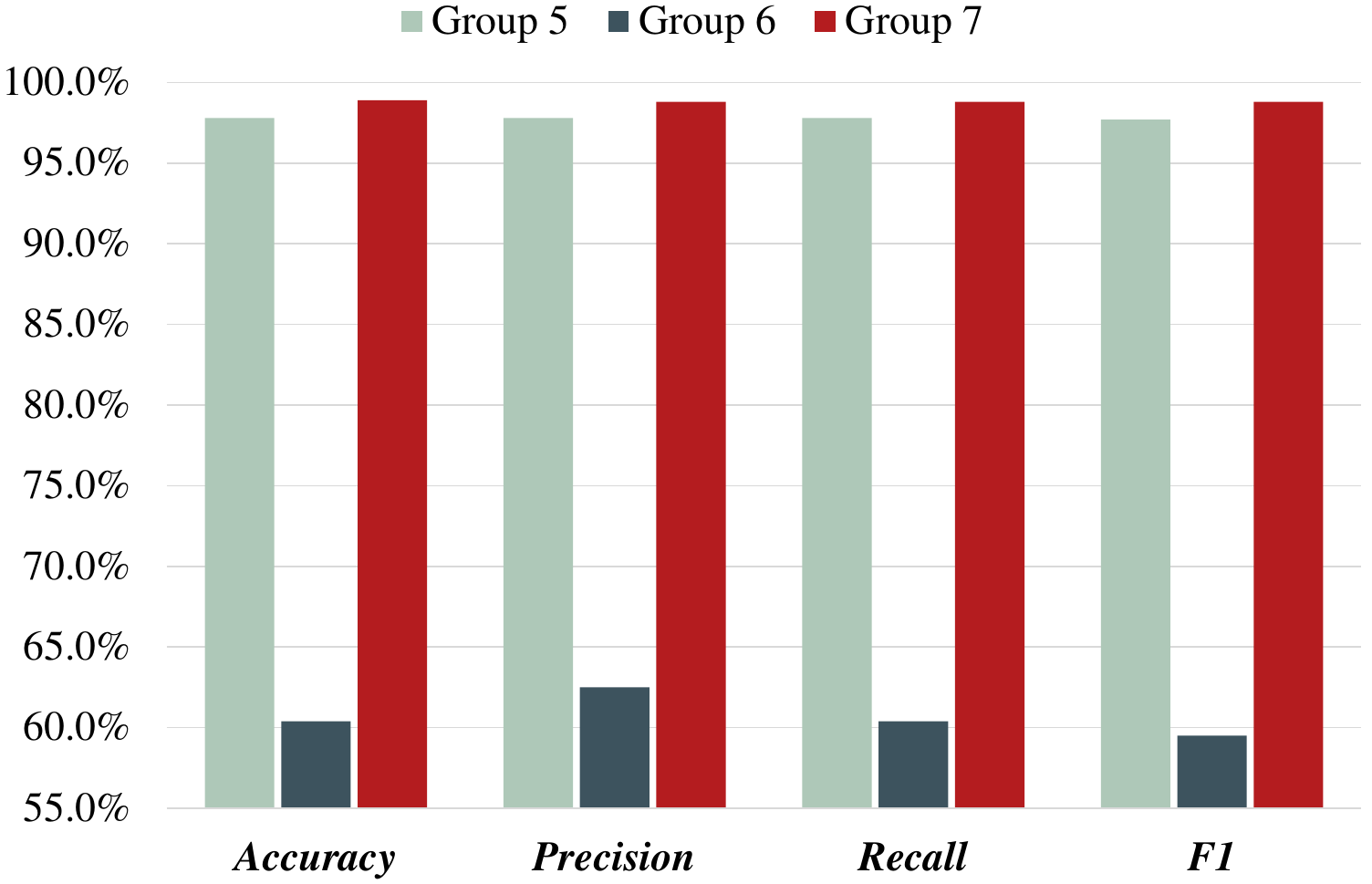}
\caption{Histogram comparing the data of double-module ablation experiments.}
\label{fig_9}
\end{figure}

From Fig. \ref{fig_9} and Table \ref{tab:table8}, it is evident that the absence of the BiLSTM block (Group 6) results in a pronounced decline in multiple performance metrics compared to other groups. This underscores the critical role of the BiLSTM block in characterizing sample-level features. Relying solely on the Transformer-encoder block to directly process frame-level features proves challenging for achieving coherent global feature integration. 

Additionally, the omission of the ConvLSTM block (Group 5) leads the Transformer-encoder to exclusively apply multi-head self-attention mechanisms to sample-level features, causing an almost complete loss of frame-level information when handling global data. Moreover, since the BiLSTM inherently addresses sample-level information, the functional overlap with the Transformer-encoder results in recognition performance almost identical to utilizing BiLSTM alone for frame-level information processing (Group 2). 

Ultimately, Group 7 attains the most outstanding recognition performance, surpassing the complete model. Through preceding analyses, it becomes apparent that, despite lacking the global information processing block of the Transformer-encoder, the ConvLSTM and BiLSTM blocks individually process raw data at both frame and sample levels. This approach of finely extracting local and global features effectively characterizes audio features.

In summary, the superior fitting of numerical audio data with temporal information by the Transformer-encoder, when coupled with effective processing of frame-level and sample-level information by the two blocks, results in improved handling of dual-level information (Group 4). This significantly reduces training time compared to the simple summation of the effects of the three modules independently. Evaluating metrics demonstrates that the proposed model in this paper outperforms any experimentally designed model from other groups in terms of comprehensiveness and accuracy. It also showcases superior performance in the weighted metric of overall performance, i.e., the F1 score, achieving a balance between evaluation metrics and training costs.

\subsection{Transfer Experiment: Deeper Exploration of Model Performance (RQ3)}\label{5.8}

\begin{table*}[h]
	\caption{Transfer experimental data.} \label{tab:table9}
	\centering
	\begin{tabular*}{0.7\linewidth}{ccccccc}
		\hline
		\multirow{2}{*}{} & \multirow{2}{*}{Pre-trained Model} & \multirow{2}{*}{Trainable\_Layer} & \multicolumn{4}{c}{Evaluation metrics} \\ \cline{4-7} 
		&                                    &                                   & Accuracy      & Precision     & Recall     & F1      \\ \hline
		Group8            & Transformer\_encoder               & MlP(128)+Softmax(21)              & 69.5\%   & 68.3\%  & 69.5\%  & 68.1\%  \\
		Group9            & Transformer\_encoder               & Softmax(21)                       & 75.0\%   & 75.3\%  & 75.0\%  & 74.4\%  \\
		Group10           & All of blocks                      & MlP(128)+Softmax(21)              & 87.9\%   & 88.0\%  & 87.9\%  & 87.7\%  \\
		Group11           & All of blocks                      & Softmax(21)                       & 86.2\%   & 86.6\%  & 86.2\%  & 85.7\%  \\ \hline
	\end{tabular*}
\end{table*}

The previous experiments validated the overall model performance and contributions of individual blocks. In this segment, we utilize the model trained on the CCNU\_Mobile dataset as a pre-trained model for fine-tuning on the MOBIPHONE dataset to evaluate the model's transfer performance. Two models are employed as pre-trained models: one retaining only the Transformer-encoder block in the backend and the other being the complete model. Transfer performance is assessed in stages. As the number of recognized classes is adjusted from 45 to 21 during transfer, adjustments are made to the softmax layer in the pre-trained model. For fine-tuning network layers, two sets of experiments are conducted: one with the last layer of the MLP having 128 nodes + softmax and the other with only softmax having 21 nodes. Transfer experiment results under the four evaluation metrics are presented in Table \ref{tab:table9}.

During fine-tuning of the pre-trained model, the learning rate and batch size are reduced, and the number of training epochs is increased to enable the target model to progressively learn details from the pre-trained model. The original parameters for the pre-trained model were a learning rate of 1e-4, batch size of 64, and 100 training epochs. During fine-tuning, these three parameters are adjusted to 1e-5, 32, and 300, respectively.

\begin{figure}[h]
\centering
\includegraphics[width=3.5in]{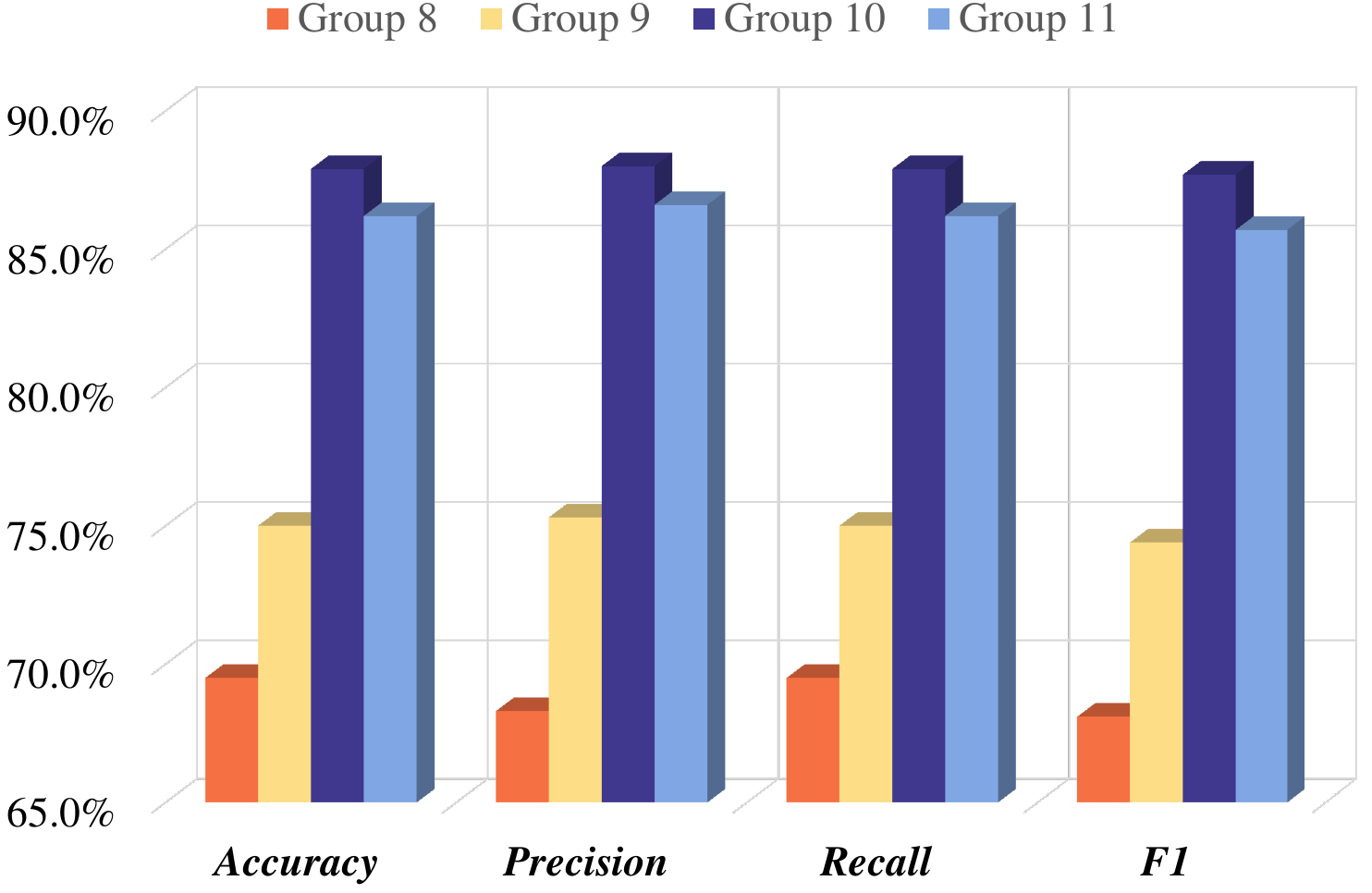}
\caption{Bar chart comparing the results of transfer learning experiments.}
\label{fig_10}
\end{figure}

From the experimental data in the first two groups in Table \ref{tab:table8}, it can be observed that when only the Transformer-encoder block is used for transfer, an accuracy of 75.0\% and an F1-Score of 74.4\% are achieved. This indicates that this module possesses a certain degree of transferability, laying the foundation for the overall model transfer. Furthermore, the performance under fine-tuning only the softmax surpasses that of simultaneously fine-tuning the MLP (128) + Softmax in the pre-trained model. This is because when the Transformer-encoder block is used as a pre-trained model, it does not jointly process frame-level and sample-level features, only performing global information fusion on the frontend features. Its representation capability is insufficient, leading to the last layer of the MLP in the pre-trained model not effectively transmitting the depth information of the frontend features to the fine-tuned classifier. Consequently, a decline in evaluation metrics occurs. It is also noteworthy that, in this specific scenario, the precision is 1.2 percentage points lower than the recall (while in other scenarios, the difference is within 0.5\% and precision is higher), indicating that the transferred model prioritizes recall performance but sacrifices some precision.

From the last two groups of experiments in Table \ref{tab:table9}, it can be seen that when the pre-trained model is the complete model proposed in this paper, the transferred performance is quite commendable, with an accuracy of 87.9\%. Moreover, fine-tuning the last layer of the pre-trained model's MLP has a positive effect on the classification results. This indicates that the proposed model in this paper effectively explores the frame-level and sample-level features of the original data, and through the Transformer-encoder block, further facilitates reasonable information interaction at a global scale. The pre-trained model efficiently transfers the well-processed feature data to the modified classifier through the last layer of the MLP, achieving effective transfer and comprehensively improving the evaluation metrics. In addition, within a reasonable range where precision and recall differ by less than 0.5\%, the F1-score has also increased by 2\%, once again confirming that the proposed model in this paper has good transferability. By transferring the model proposed in this paper, the new model finds a reasonable balance between recall and precision, validating that introducing the last layer of the MLP from the pre-trained model into training has a positive promoting effect on transfer learning.

\section{Conclusion} \label{Con}

This paper proposes a method for recognizing mobile recording devices, employing modeling at both the frame-level and sample-level scales. The method integrates two main components: frontend feature extraction and backend recognition model construction. In frontend feature extraction, features are extracted at multiple-shallow, and the temporally aligned features are concatenated to comprehensively represent the original audio. In the construction of the backend model, the paper integrates features fully from the frame-level, sample-level, and cross-scale for effective decision-making. Specifically, it enhances frame-level features using one-dimensional ConvLSTM, processing short-term discrete frequency information at the frame-level feature scale. Subsequently, it conducts staged long-term analysis at the sample-level scale on the temporally enhanced two-dimensional features. Finally, features are fused at the cross-scale. The experiments also confirm that this model exhibits robust transfer performance. However, there are some limitations in this study, such as the discussion on the potential to further streamline model parameters for a significant reduction in training time being less comprehensive. Additionally, addressing how to transfer to a test set with a distribution different from the two datasets used in the transfer experiments requires further consideration.

\section*{Abbreviations}
The following abbreviations are used in this manuscript:

\begin{table}[h]
\centering
\begin{tabular}{ll}
MFCC     & Mel-Frequency Cepstral Coefficients                     \\
ConvLSTM & Convolutional Long Short-Term Memory                    \\
BiLSTM   & Bidirectional Long Short-Term Memory                    \\
BED      & Band Energy Difference                                  \\
VAD      & Voice Activity Detection                                \\
GMM      & Gaussian Mixture Models                                 \\
GSV      & Gaussian Supervector Vectors                            \\
SVM      & Support Vector Machine                                  \\
BERT     & Bidirectional Encoder Representations from Transformers \\
RSF      & Random Spectral Features                                \\
LSF      & Labeled Spectral Features                               \\
VAD      & Voice Activity Detection                                \\
LFCC     & Linear Frequency Cepstral Coefficients                  \\
LPCC     & Linear Prediction Cepstral Coefficients                 \\
CMVN     & Cepstral Mean and Variance Normalization                \\
SRC      & Sparse Representation Classifier                        \\
VQ       & Vector Quantization                                     \\
RBF-NN   & Radial Basis Function-Neural Network                    \\
DFT      & Discrete Fourier Transform                              \\
RD-TCNs      & Residual Dense Time Convolutional Networks                            \\
CNN      & Convolutional Neural Network                            \\
ResNet   & Residual Neural Network                                 \\
LSTM     & Long Short-Term Memory                                  \\
FFT      & Fast Fourier Transform                                  \\
DCT      & Discrete Cosine Transform                               \\
LN       & Layer Normalization                                     \\
MLP      & MultiLayer Perceptron                                   \\
Adam     & Adaptive Moment Estimation                              \\
TIMIT    & Texas Instruments/Massachusetts Institute of Technology \\
DNN      & Deep Neural Network                                    
\end{tabular}
\end{table}
\nocite{*}



\bibliography{240726IEEECE,Citations}
\bibliographystyle{IEEEtran}
\end{document}